\documentclass[12pt]{iopart}

\usepackage[dvips]{graphicx}

\newcommand{\real}{\mathcal{R}e}
\newcommand{\imag}{\mathcal{I}m}

\newcommand{\G}{\mathcal{G}}

\newcommand{\C}{\mathcal{C}}

\begin{document}

\title[Parallel algorithm with spectral convergence]
       {Parallel algorithm with spectral convergence
       for nonlinear integro-differential equations}

\author{Bogdan Mihaila\dag}
\address{\dag\
Physics Division, Argonne National Laboratory, Argonne, IL 60439}
\ead{bogdan@theory.phy.anl.gov}

\author{Ruth E. Shaw\ddag}
\address{\ddag\
Department of Applied Statistics and Computer Science,
University of New Brunswick, Saint John, NB Canada E2L 4L5}
\ead{reshaw@unbsj.ca}

\begin{abstract}
We discuss a numerical algorithm for solving nonlinear
integro-differential equations, and illustrate our findings for
the particular case of Volterra type equations. The algorithm
combines a perturbation approach meant to render a linearized
version of the problem and a spectral method where unknown
functions are expanded in terms of Chebyshev polynomials
(El-gendi's method).
This approach is shown to be suitable for the calculation of
two-point Green functions required in next to leading order studies of
time-dependent quantum field theory.
\end{abstract}


\submitto{\JPA}

\maketitle

%
%

\section{Introduction}
\label{sec:intro}

Astrophysical applications related to the physics of the early
universe, as well as challenges posed by the physics programs at
new heavy ion accelerators, have triggered a renewed interest in
the understanding of real time processes in the context of quantum
field theory. With the advent of new computer technology and the
recent success of new computational schemes, non-equilibrium
phenomena which have been previously studied only in the framework
mean-field theory~\cite{ref:Hartree,ref:GV,ref:LOLN}, are now
being revisited, and more complex next to leading order
approaches~\cite{ref:EQT,ref:abw,ref:ctpN,ref:berges} are being
used in an attempt to clarify the role played by the rescattering
mechanism, which is responsible for driving an out of equilibrium
system back to equilibrium. Of particular interest is the study of
the dynamics of phase transitions and particle production
following a relativistic heavy-ion collision. One way of
approaching this study is based on solving Schwinger Dyson
equations within the closed time path (CTP)
formulation~\cite{ref:SBMK}. This formalism has been recently
shown to provide good approximations of the real time evolution of
the system both in quantum mechanics and 1+1 dimensional classical
field theory~\cite{ref:MDC}, where direct comparisons with exact
calculations can be performed.

The key element in carrying out such studies is related to the
calculation of the two-point Green function, which is solved for
self-consistently with the equations of motion for the fields. The
two-point Green function gives rise to Volterra-type integral or
integro-differential equations. In the process of extending our
study to encompass a higher number of spatial dimensions, i.e. 2+1
and 3+1 field theory, we are faced with the challenge of coping
with constraints dictated both by storage and time-related
computational limits. Thus our interest in designing algorithms
which feature spectral convergence in order to achieve convergence
with minimum storage requirements. In addition, we also desire
these algorithms to scale when ported to massively multiprocessor
(MPP) machines, so that solutions can be obtained in a reasonable
amount of time.

Algorithms for Volterra integral and integro-differential
equations usually start out at the lower end of the domain, $a$,
and march out from  $x = a$, building up the solution as they
go~\cite{ref:nr}. Such methods are \emph{serial} by nature, and
are, in general, not suitable for parallel implementation on a MPP
machine. Even so, clever approaches to already existing methods
can provide algorithms that take advantage of a parallel
processing computer: Shaw~\cite{ref:shaw} has shown recently that
once the starting values of the approximation are obtained, one
can design a \emph{global} approach where successive
approximations of the solution over the entire domain $x \in
[a,b]$ can be evaluated simultaneously.

In a recent paper~\cite{ref:bm} one of us has discussed a spectral
method~\cite{ref:el_gendi} of solving some types of equations of
interest for the study of time-dependent nonequilibrium problems
in quantum field theory. The gist of the method consists in
expanding out the unknown function in terms of Chebyshev
polynomials on a suitable grid, thus reducing the problem to
finding the numerical solution of a system of linear equations.
The main advantage of this method over standard finite-difference
type methods resides in the spectral character of its convergence.
This is related in part to the fact that Chebyshev type methods
use a non-uniform grid, while finite-difference methods require a
uniform grid. Usually there is a trade-off between computational
time and storage requirements, and a balanced solution must be
reached on a case-by-case basis. Spectral methods are more
expensive per point as the matrices may be considerably denser
than in the finite-difference case, but we require considerably
fewer grid points in order to achieve the same degree of accuracy.
By expanding the unknown function on a compact support in
Chebyshev polynomials and using a partition of the domain based
either on the set of (N+1) extrema or the set of~N zeros of
$T_N(x)$ -- the Chebyshev polynomial of first kind of degree~N --
we in fact replace a continuous problem by a discrete one. For
non-singular functions the discrete orthogonality and completeness
relations for Chebyshev polynomials at the above grid points
assure a \emph{defacto} exact expansion for an \emph{arbitrary}
finite value $N$. In practice however, one has to compute
derivatives and integrals of the unknown function at the
collocation points, and the Chebyshev expansion provides only an
approximation for these subsequent computations. These errors,
together with the finite accuracy of numerical methods needed in
conjunction with the Chebyshev expansion, conspire in order to
deteriorate the accuracy of the solution at very small values
of~N.

The paper is organized as follows: In Section~\ref{sec:finite},
for comparison purposes, we start by reviewing a finite-difference
approach for the numerical solution of Volterra type
integro-differential equations. We review the general framework of
the Chebyshev-expansion method in Section~\ref{sec:cheby}, and
illustrate our approach 
for the case of Volterra integro-differential equations. In
Section~\ref{sec:test} we present a complete assessment of the
convergence and computational cost of the proposed method for the
case of a test problem, and compare with results obtained via the
finite-difference method. In Section~\ref{sec:ctp} we discuss the
relevant aspects of a large-scale calculation arising in the study
of time-dependent quantum field theory, for which our numerical
strategy is particularly suitable. We present our conclusions in
Section~\ref{sec:concl}.

%
%

\section{Stable multi-step method for Volterra type equations}
\label{sec:finite}

The type of problems arising in the study of time-dependent
nonequilibrium quantum field theory via a Schwinger-Dyson equation
approach, can be formally reduced to the general case of a
nonlinear Volterra integro-differential equation. Direct methods
for solving nonlinear Volterra integral and integro-differential
equations are inherently serial and therefore have not received
much attention for use on a parallel computer. It is worth
mentioning here the work of Crisci \emph{et al}~\cite{ref:VRK},
who concentrated on the stability aspects of parallel iteration of
Volterra-Runge-Kutta (VRK) methods for solving Volterra integral
equations on parallel computers. VRK methods are step-by-step
methods and can take advantage of parallel architecture. Sommeijer
\emph{et al}~\cite{ref:SCH} covered the stability of parallel
block methods for ordinary differential equations (ODE) and
included equations of the integro-differential type in their
discussion.

We summarize here a recent parallel algorithm~\cite{ref:shaw},
which concentrates on modifying the algorithmic side of the
numerical solution process for use on a parallel processor while
consciously utilizing methods that are known to be stable. The
algorithm is in effect an example of a higher-order
finite-difference approach, and we use this approach to compare
with the spectral method presented later in this paper.


For illustration, let us consider a first-order nonlinear Volterra
integro-differential equation of the form
\begin{eqnarray}
   &&
   \mathbf{y}'(x) \ = \ F[x,\mathbf{y},Z[x;\mathbf{y}]]
   \>,
   \quad x \in [a,b] \>,
\label{eq:ode2_nlin}
\end{eqnarray}
with
\begin{eqnarray}
   Z[x;\mathbf{y}] = \int_{a}^x K[x,t;\mathbf{y}(t)] \mathrm{d}t
   \>,
\label{eq:zz}
\end{eqnarray}
and subject to the initial condition
\begin{eqnarray}
   \mathbf{y}(a) = y_0
   \>.
\label{eq:bv_cond}
\end{eqnarray}
Let $I_N$ be a partition of I=[a,b], where $I_N=\{x_N=a + nh,\
n=0(1)\,N,\ Nh=(b-a)\}$. The problem is to find approximations
$y_n$ to the solution $\mathbf{y}(x_n)$ of
Eqs.~(\ref{eq:ode2_nlin}--\ref{eq:bv_cond}) for each $x_n \in
I_N$. A $k$-step method for an integro-differential equation of
the form~(\ref{eq:ode2_nlin}) is given by
\begin{eqnarray}
   y_{n+1} = y_n + h \sum_{j=0}^k w_j F(x_{n-j},y_{n-j},z_{n-j})
   \>, \quad n=k(1)\,N \>,
\end{eqnarray}
where
\begin{eqnarray}
   z_{n-j} = h \sum_{i=0}^{n-j} c_{n-j,i} K(x_{n-j},x_i,y_i) \>,
   \quad j = 0(1)\,k \>,
   \quad
   y_0 = \mathbf{y}(a)
   \>.
\end{eqnarray}
The weights $w_i$ depend on the $k$-step method selected and the
weights $c_{i,j}$ are those of a standard quadrature formula for
integrating a function whose value is known at equally spaced
steps, such as a Newton-Cotes or Newton-Gregory quadrature rule.
For our multi-step ($k=4$) method~\cite{ref:nr} we choose the
fourth order Adams-Bashforth predictor
\begin{eqnarray}
   y^{0}_{k+1} &= y_k + \frac{h}{24}
   \Bigl [ &55\, F(x_k, y_k, z_k)
         - 59\, F(x_{k-1}, y_{k-1}, z_{k-1})
   \\ \nonumber &&
         + 37\, F(x_{k-2}, y_{k-2}, z_{k-2})
         - 9\,  F(x_{k-3}, y_{k-3}, z_{k-3})
   \Bigr ] \>,
\end{eqnarray}
and the Adams-Moulton corrector
\begin{eqnarray}
   y_{k+1} &= y_k + \frac{h}{24}
   \Bigl [ &9\, F(x_{k+1}, y^{0}_{k+1}, z_{k+1})
         + 19\, F(x_k, y_k, z_k)
   \\ \nonumber &&
         -  5\, F(x_{k-1}, y_{k-1}, z_{k-1})
         +    F(x_{k-2}, y_{k-2}, z_{k-2})
   \Bigr ] \>,
\end{eqnarray}
while the integral term~(\ref{eq:zz}) is calculated based on the
Newton-Gregory quadrature formula. We use a fourth order
Runge-Kutta method in order to start out the calculation.

In order to make the algorithm suitable for parallel processing,
it is useful to recall that a standard quadrature method based on
an uniform grid for the integral term~$z_i$ requires knowledge of
the integrand function at the abscissas in the interval
$[x_0,x_i]$. This is obviously a serial process and not a good
candidate for parallelization. It can be observed however, that
once the starting values are obtained, \emph{all} approximations
$z_i$ with $i=0(1)\,k-1$ can simultaneously be evaluated up to and
including $x_{k-1}$. After that, once a value of $y_j$
corresponding to a new step $x_j$ is established via the
predictor-corrector method, all values $z_i$ with $i=j(1)\,N$ can
also be evaluated simultaneously. This observation makes the
following algorithm possible:
\begin{enumerate}




   \item Find the starting values $(y_i,z_i)$ with $i=0(1)\,k-1$

   \item $do\  i = k,N$

    \hspace{0.2in}add contributions to $z_i$ corresponding to $(x_j,y_j)$, where $j=0(1)\,k-1$

   \item 
   $do\  i = k,N$
      \begin{enumerate}
         \item predict $y_{i}$
         \item estimate $z_{i}$ from $(x_{i},y_{i})$
         \item correct $y_{i}$
         \item $do\  j = i,N$

               \hspace{0.2in}update $z_j$ by adding the contribution corresponding to
               $(x_{i},y_{i})$
      \end{enumerate}
\end{enumerate}

The above numerical algorithm is implemented using the OpenMP
style directives for the Portland Group's pgf77 FORTRAN compiler,
and reportedly shows good scalability on a shared-memory
multiprocessor. The speedup of the finite difference method is
best for a large number of grid points which, correspondingly,
gives a better solution approximation. For example, with N=5120
and 4 processors the speedup is 3.86,
a good measure of processor utilization.

While the preceding algorithm performs well on a shared memory
platform, it does not port easily to an MPP machine. Before we
comment on the efficiency of the algorithm, let us make two
general comments: Firstly, we denote by $T_{\textrm {calc}}$ and
$T_{\textrm {comm}}$ the time required to perform a floating-point
operation and the time required to send a floating-point number,
respectively. Secondly, we will ignore for simplicity the effect
of message sizes on communication costs, and assume throughout
that the ratio $T_{\textrm {comm}}/T_{\textrm {calc}}$ is
independent of $N$.

Returning now, to our proposed algorithm, we remark that the
communication cost for the corresponding implementation involves
only the integral terms. Even so, using the message-passing
interface (MPI) protocol 
the communication cost is $4\log N$ for the starting values and up
to $N^2$ for the remainder of the algorithm which gives a total of
$(N^2 + 4\log N) T_{\textrm {comm}}$. The total number of flops
depends on the specific application but a reasonable measure is
the number of function evaluations which is given by $(N^2 + 4N)
T_{\textrm {calc}}$. The ratio of communication to computation
\[ \frac{N^2 + 4\log N}{N^2 + 4N} \frac{T_{\textrm {comm}}}{T_{\textrm {calc}}}
\]
approaches a \emph{constant} value as $N$ gets larger. The
communication overhead problem can be relaxed by employing a
spectral method discussed in the following section, the
improvement being especially significant for a multidimensional
problem of the type required by our nonequilibrium quantum field
theory calculations~\cite{ref:MDC}.

%
%

\section{Spectral method with Chebyshev polynomials}
\label{sec:cheby}

Consider the $N+1$ extrema of the Chebyshev polynomial of the
first kind of degree~$N$, $T_N(x)$. This set defines a non-uniform
grid in the interval $[-1, 1]$, as
\begin{equation}
   \tilde{x}_k \ = \
   \cos \left ( \frac{\pi k}{n} \right ) \>,
   \quad k = 0(1)\,N
   \>.
   \label{eq:Tn_max}
\end{equation}
On this grid, the Chebyshev polynomials of degree $i<n$ obey
discrete orthogonality relations
            \begin{equation}
               \sum_{k=0}^N {\rm {}''} \ T_i(\tilde{x}_k) T_j(\tilde{x}_k)
               \ = \
               \beta_i \ \delta_{i \, j}
               \>,
            \label{eq:cheby_ortog_b}
            \end{equation}
            where the constants $\beta_i$ are
            \[
               \beta_i
               \ = \
               \left \lbrace
                     \begin{array}{ll}
                        \displaystyle{\frac{N}{2}}  \>, & i \neq 0,N \>,\\
                        N                           \>, & i = 0,N \>.
                     \end{array}
               \right .
            \]
Here, the summation symbol with double primes denotes a sum with
both the first and last terms halved. We approximate an arbitrary
continuous function of bounded variation $f(x)$ in the interval
$[-1,1]$, as
\begin{eqnarray}
   f(x)
   & \approx &
   \sum_{j=0}^{N} {\rm {}''} \ b_j \ T_j(x)
   \>,
\label{eq:f_approx_b}
\end{eqnarray}
with
\begin{eqnarray}
   b_j
   & = &
   \frac{2}{N} \,
   \sum_{k=0}^N {\rm {}''} \
       f(\tilde{x}_k) T_j(\tilde{x}_k) \>, \quad j = 0(1)\,N
   \>.
\label{eq:coeff_b}
\end{eqnarray}
Eq.~(\ref{eq:f_approx_b}) is exact at {\em x} equal to $\tilde
x_k$ given by Eq.~(\ref{eq:Tn_max}). Based on
Eq.~(\ref{eq:f_approx_b}), we can also approximate derivatives and
integrals as
\begin{eqnarray}
   f'(x)
   & \approx &
   \sum_{k=0}^N {\rm {}''} \
       f(\tilde{x}_k)  \
   \frac{2}{N} \,
   \sum_{j=0}^{N} {\rm {}''} \
       T_j(\tilde{x}_k) \ T_j'(x)
   \>.
\label{eq:f_derivative_b}
\end{eqnarray}
and
\begin{eqnarray}
   \int_{-1}^{x} f(t) \, \mathrm{d}t
   & \approx &
   \sum_{k=0}^N {\rm {}''} \
       f(\tilde{x}_k)  \
   \frac{2}{N} \,
   \sum_{j=0}^{N} {\rm {}''} \
       T_j(\tilde{x}_k) \ \int_{-1}^{x} T_j(t) \, \mathrm{d}t
   \>.
\label{eq:f_int_b}
\end{eqnarray}
In matrix format, we have
\begin{eqnarray}
   \left [ \int_{-1}^x \ f(t) \, \mathrm{d}t \right ]
   & \approx &
   \tilde S \ \left [ f \right ]
   \>,
\label{eq:Beqn_b}
   \\
   \left [ f'(x) \right ]
   & \approx &
   \tilde D \ \left [ f \right ]
   \>,
\label{eq:BTeqn_b}
\end{eqnarray}
The elements of the column matrix $\left [ f \right ]$ are given
by $f(\tilde x_k), \ k=0(1)\,N$. The right-hand side of
Eqs.~(\ref{eq:Beqn_b}) and~(\ref{eq:BTeqn_b}) give the values of
the integral~$\int_{-1}^x \ f(t) \, \mathrm{d}t$ and the
derivative~$f'(x)$ at the corresponding grid points, respectively.
The actual values of the elements of the matrices $\tilde S$ and
$\tilde D$ can be derived using Eqs.~(\ref{eq:f_derivative_b},
\ref{eq:f_int_b}).

%
%


In order to illustrate the Chebyshev algorithm, we consider again
the case of a first-order nonlinear Volterra
integro-differential equation of the form
\begin{eqnarray*}
   &&
   \mathbf{y}'(x) \ = \ F[x,\mathbf{y},Z[x;\mathbf{y}]]
   \>,
   \quad x \in [a,b] \>,
   \\ &&
   Z[x;\mathbf{y}] = \int_{a}^x K[x,t;\mathbf{y}(t)] \mathrm{d}t
   \>,
\end{eqnarray*}
with the initial condition
\begin{eqnarray*}
   \mathbf{y}(a) = y_0
   \>.
\end{eqnarray*}
Here we make no explicit restrictions on the actual form of the
function $F[x,\mathbf{y},Z[x;\mathbf{y}]]$, so both linear and
nonlinear equations are included. We determine the unknown
function $\mathbf{y}(x)$ using a perturbation approach: We start
with an initial guess of the solution $\mathbf{y}_0(x)$ that
satisfies the initial condition $\mathbf{y}_0(a) = y_0$, and write
\[
   \mathbf{y}(x) \ = \ \mathbf{y}_0(x) \ + \ \epsilon(x)
   \>,
\]
with $\epsilon(x)$ being a variation obeying the  initial
condition
\begin{equation}
   \epsilon(a) = 0 \>.
\label{eq:initial}
\end{equation}
Hence, the original problem reduces to finding the perturbation
$\epsilon(x)$, and improving the initial guess in a iterative
fashion.

We use the Taylor expansion of $F[x,\mathbf{y},Z[x;\mathbf{y}]]$
about $\mathbf{y}(x)=\mathbf{y}_0(x)$ and keep only the linear
terms in $\epsilon(x)$ to obtain an equation for the variation
$\epsilon(x)$
\begin{eqnarray}
   \epsilon'(x)
   &-&
   \left . \frac{\partial F[x,\mathbf{y},Z[x;\mathbf{y}]]}{\partial \mathbf{y}(x)}
   \right |_{\mathbf{y}(x)=\mathbf{y}_0(x)}
   \epsilon(x)
   \nonumber \\
   &-&
   \left . \frac{\partial F[x,\mathbf{y},Z[x;\mathbf{y}]]}{\partial Z[x;\mathbf{y}]}
   \right |_{\mathbf{y}(x)=\mathbf{y}_0(x)}
   \int_{a}^x
       \left . \frac{\partial K[x,t;\mathbf{y}(t)]}{\partial \mathbf{y}(x)}
       \right |_{\mathbf{y}(x)=\mathbf{y}_0(x)}
       \epsilon(t) \mathrm{d}t
   \nonumber \\ &&
   =
   - \mathbf{y}_0'(x) + F[x,\mathbf{y}_0(x),Z[x;\mathbf{y}_0(x)]]
   \>.
\label{eq:eps_eqn}
\end{eqnarray}
Equation~(\ref{eq:eps_eqn}) is of the general form (\ref{eq:ode2})
\begin{equation}
   \epsilon'(x) = q[x,\epsilon(x)] + r(x)
   \> ,
\label{eq:ode2}
\end{equation}
where
\begin{eqnarray*}
   q[x,\epsilon(x)]
   =
   \left . \frac{\partial F[x,\mathbf{y},Z[x;\mathbf{y}]]}{\partial \mathbf{y}(x)}
   \right |_{\mathbf{y}(x)=\mathbf{y}_0(x)} \epsilon(x)
   \\ \qquad
   +
   \left . \frac{\partial F[x,\mathbf{y},Z[x;\mathbf{y}]]}{\partial Z[x;\mathbf{y}]}
   \right |_{\mathbf{y}(x)=\mathbf{y}_0(x)}
   \int_{a}^x
       \left . \frac{\partial K[x,t;\mathbf{y}(t)]}{\partial \mathbf{y}(x)}
       \right |_{\mathbf{y}(x)=\mathbf{y}_0(x)}
       \epsilon(t) \mathrm{d}t
   \>,
\end{eqnarray*}
and
\begin{eqnarray*}
   r(x)
   & = &
   - \mathbf{y}_0'(x) - F[x,\mathbf{y}_0(x),Z[x;\mathbf{y}]]
   \>,
\end{eqnarray*}
together with the initial condition given by~(\ref{eq:initial}).
We replace Eqs.~(\ref{eq:ode2}) and~(\ref{eq:initial}) by an
integral equation, obtained by integrating Eq.~(\ref{eq:ode2}) and
using the initial condition~(\ref{eq:initial}) to choose the lower
bound of the integral. We obtain
\begin{eqnarray}
\label{eq:int_ic}
   \epsilon(x)
   =
   \int_{a}^x q[t,\epsilon(t)] \ \mathrm{d}t
   +
   \int_{a}^x r(t) \mathrm{d}t
   \>,
\end{eqnarray}
which is in fact a \emph{linear} Volterra integral equation of the
second kind. Using the techniques developed in the previous
section to calculate integrals, the integral
equation~(\ref{eq:int_ic}) can be transformed into a linear system
of equations. A practical implementation of this algorithm is
illustrated via a test problem in the following section.

%
%

\section{Test problem}
\label{sec:test}

Following Shaw~\cite{ref:shaw}, we consider the test problem
\begin{eqnarray}
   &&
   \mathbf{y}(x) = x e^{1 -\mathbf{y}(x)} - \frac{1}{(1+x)^2} - x - \int_0^x
   \frac{x}{(1+t)^2} e^{1 -\mathbf{y}(t)} \mathrm{d}t \>,
   \\ &&
   \mathbf{y}(0) = y_0 = 1 \>, \qquad x \in [0,1] \>,
\end{eqnarray}
which has the exact solution
\begin{eqnarray}
   \mathbf{y}(x) = \frac{1}{1+x} \>.
\end{eqnarray}
We shall use the initial guess $\mathbf{y}_0(x) = y_0 \cos(x)$, so
that $\mathbf{y}_0(0) = y_0$. The equation for the variation
$\epsilon(x)$ is
\begin{eqnarray}
   &&
   \epsilon(x)
   - \int_0^x t e^{1 - \mathbf{y}_0(t)} \epsilon(t) \mathrm{d}t
   + \int_0^x \mathrm{d}s \int_0^s \frac{s e^{1 -\mathbf{y}_0(t)}}{(1+t)^2} \epsilon(t) \mathrm{d}t
   =
   \\ && \nonumber
   - \mathbf{y}_0(x) + y_0
   + \int_0^x \left [ t e^{1 -\mathbf{y}_0(t)} - \frac{1}{(1+t)^2} - t \right ] \mathrm{d}t
   - \int_0^x \mathrm{d}s \int_0^s
   \frac{s e^{1 -\mathbf{y}_0(t)}}{(1+t)^2} \mathrm{d}t
   \>.
\end{eqnarray}
In matrix format and using the Chebyshev expansion presented
above, the variation $\epsilon(x)$ will be obtained as the
solution of linear system of equations
\begin{equation}
   A \ \left [ \epsilon \right ] \ = \ C
   \>,
\label{eq:sislin}
\end{equation}
with matrices $A$ and $C$ given as
\begin{eqnarray*}
   A_{i \, j} & = &
   \delta_{i \, j}
   -
   \tilde S_{i \, j}
   \left [ t e^{1 - \mathbf{y}_0(t)} \right ]_j
   \nonumber \\ &&
   +
   \tilde S_{i \, k} \tilde x_k \tilde S_{k \, j} \left [ \frac{e^{1 -\mathbf{y}_0(t)}}{(1+t)^2} \right ]_j
   \>,
   \qquad
   i,j = 0(1)\,N
   \>,
   \nonumber \\
   C_i & = &
   - [ \mathbf{y}_0(t) ]_i + y_0
   + \tilde S_{i \, k}
   \left [ t e^{1 - \mathbf{y}_0(t)} - \frac{1}{(1+t)^2} - t
   \right ]_k
   \nonumber \\ &&
   -
   \tilde S_{i \, k} \tilde x_k \tilde S_{k \, \ell}
          \left [ \frac{e^{1 -\mathbf{y}_0(t)}}{(1+t)^2} \right ]_\ell
   \>.
\end{eqnarray*}

From a computational point of view the computer time is spent
initializing the matrix elements $A_{ij}$ and $C_j$ on one hand,
and finding the solution of (\ref{eq:sislin}) on the other. On the
first matter, the calculation decouples nicely, and once we have
the vector $[y_0]$, we can calculate $\{ C_i, A_{ij}, j=0(1)\,N
\}$ in parallel for $i=0(1)\,N$. The algorithm is as follows:
\begin{enumerate}
   \item calculate $[y_0] = [y_0] + [\epsilon]$ ;
   \item broadcast $[y_0]$ ;
   \item \textit{do} $i=0,N$ :
   \begin{enumerate}
      \item master to slave: send $i$ ;
      \item slave: compute $\{ C_i, A_{ij}, j=0(1)\,N \}$ ;
      \item slave to master: return $\{ C_i, A_{ij}, j=0(1)\,N \}$.
   \end{enumerate}
\end{enumerate}

Regarding the second step, i.e. solving the linear system of
equations, the best choice is to use the machine specific
subroutines, which generally outperform hand-coded solutions. When
such subroutines are not available, as in the case of a Linux
based PC cluster for instance, one can use one of the MPI
implementations available on the market. We shall see that the
efficiency of the equation solver is critical to the success of
the parallel implementation of the Chebyshev-expansion approach.
In order to illustrate this aspect we perform two calculations,
first using a LU factorization algorithm, and secondly using an
iterative biconjugate gradient algorithm. These are standard
algorithms~\cite{ref:nr} for solving systems of linear equations,
but their impact on the general efficiency of the approach is
quite different.


\subsection{Serial case}

\begin{figure}[h!]
   \centering
   \includegraphics[width=4.5in]{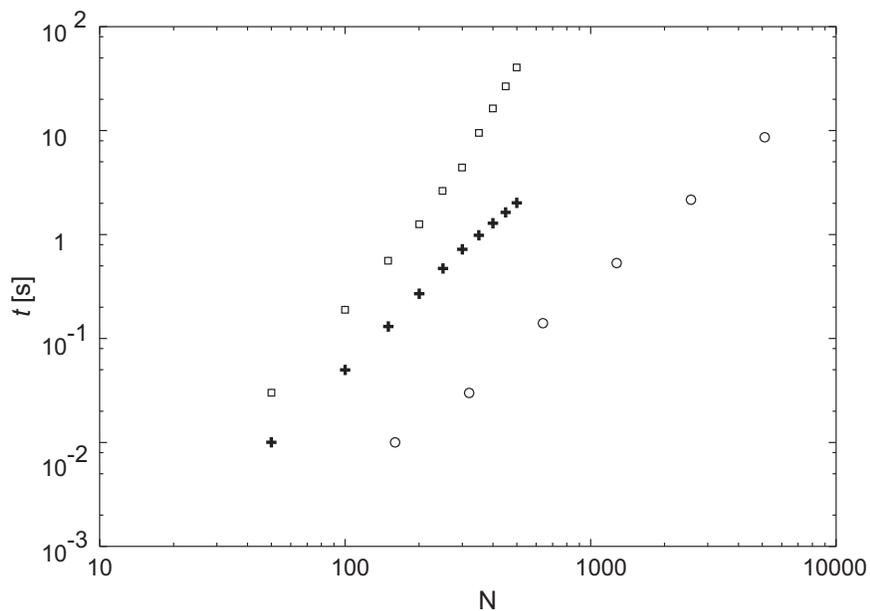}
   \caption{Average CPU time versus the number of grid points
           for the Chebyshev expansion approach using the either LU decomposition
           (squares) or the biconjugate gradient method (crosses),
           and finite-difference approach (circles).}
   \label{fig:time}
\end{figure}

\begin{figure}[h!]
   \centering
   \includegraphics[width=4.5in]{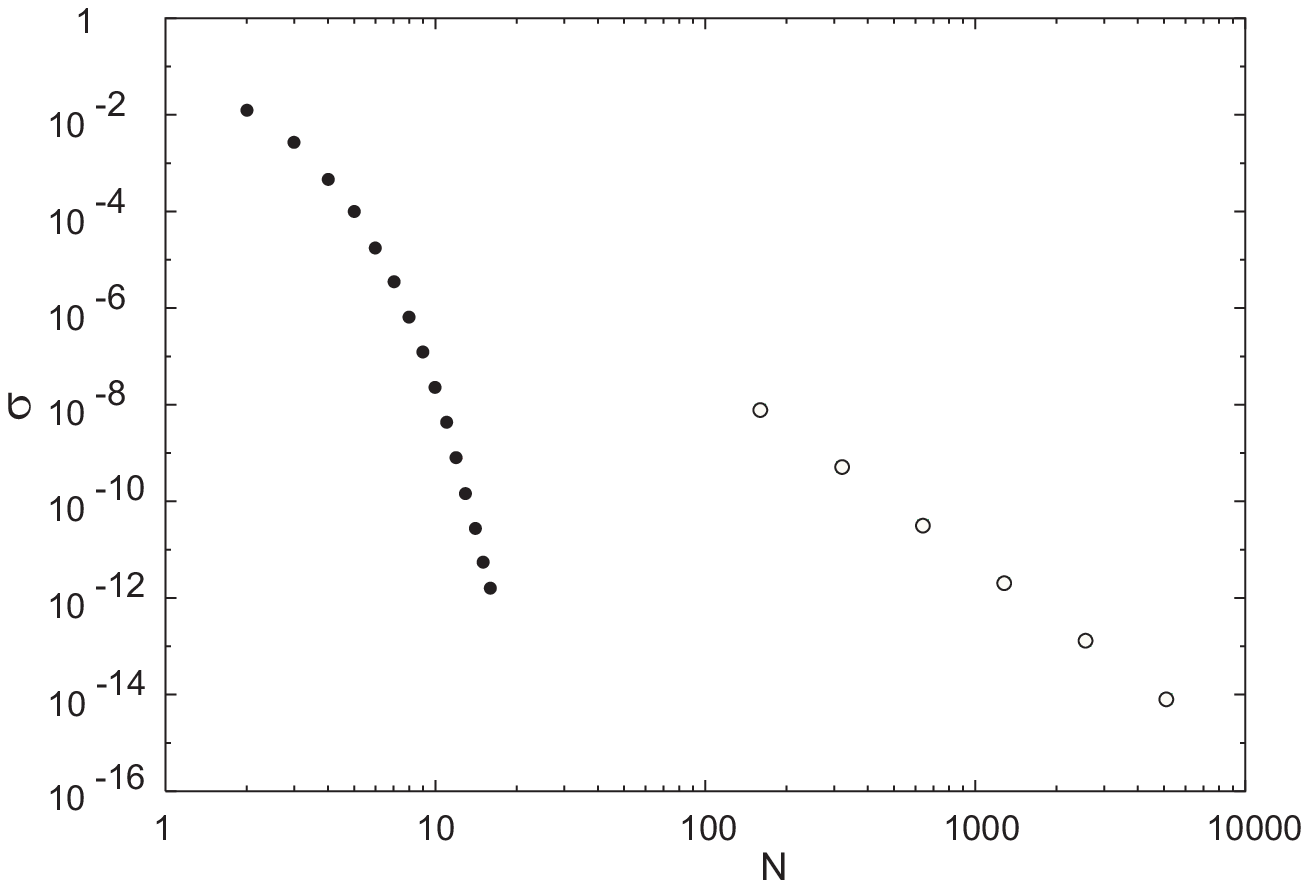}
   \caption{Convergence of the Chebyshev result (filled)
            compared with the finite-difference result (empty),
            versus the number of grid points.}
   \label{fig:conv}
\end{figure}

Figure~\ref{fig:time} depicts the average CPU time required to
complete the calculation for the various methods. 
Figure~\ref{fig:conv} illustrates the convergence of the two
numerical methods. The spectral character of the method based on
Chebyshev polynomials allows for an excellent representation of
the solution for $N>12$. We base our findings on a $\sigma <
10^{-10}$ criteria, where $\sigma$ denotes the sum of all absolute
departures of the calculated values from the exact ones, at the
grid points.

\begin{figure}[h!]
   \centering
   \includegraphics[width=4.5in]{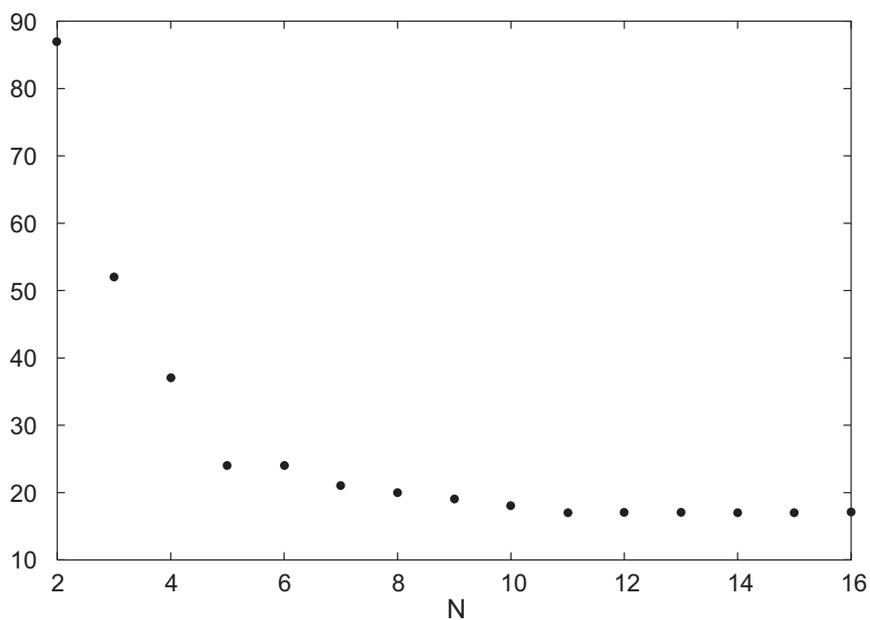}
   \caption{Number of iterations versus the number of grid points
            for the Chebyshev method.}
   \label{fig:iter}
\end{figure}

The number of iterations required to achieve the desired accuracy
in the Chebyshev case is depicted in Fig.~\ref{fig:iter}. The
number of iterations becomes flat for $N>12$, and stays constant
(17 iterations) even for very large values of N. The higher number
of iterations corresponding to the lower values of N, represents
an indication of a insufficient number of Chebyshev grid points:
the exact solution cannot be accurately represented as polynomial
of degree N for $x\in [0,1]$. It is interesting to note that for
$N=12-16$, a reasonable lower domain for the representation of the
solution using Chebyshev polynomials, the reported CPU time is so
small that for our test problem there is no real justification for
porting the algorithm to a MPP machine. This situation will change
for multi-dimensional problems such as those encountered in our
nonequilibrium quantum field theory studies.

\subsection{Parallel case}

\begin{figure}[h!]
   \centering
   \includegraphics[width=4.5in]{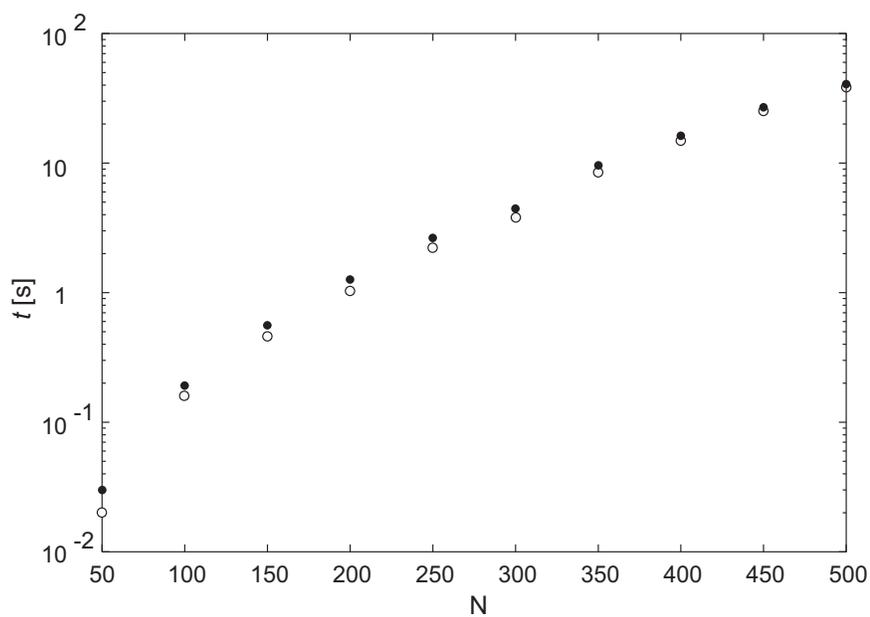}
   \caption{Total CPU time (filled) and CPU time spent carrying out
            the LU decomposition (empty), versus the number of grid points
             (1 CPU case).}
   \label{fig:time_LU}
\end{figure}

\begin{figure}[h!]
   \centering
   \includegraphics[width=4.5in]{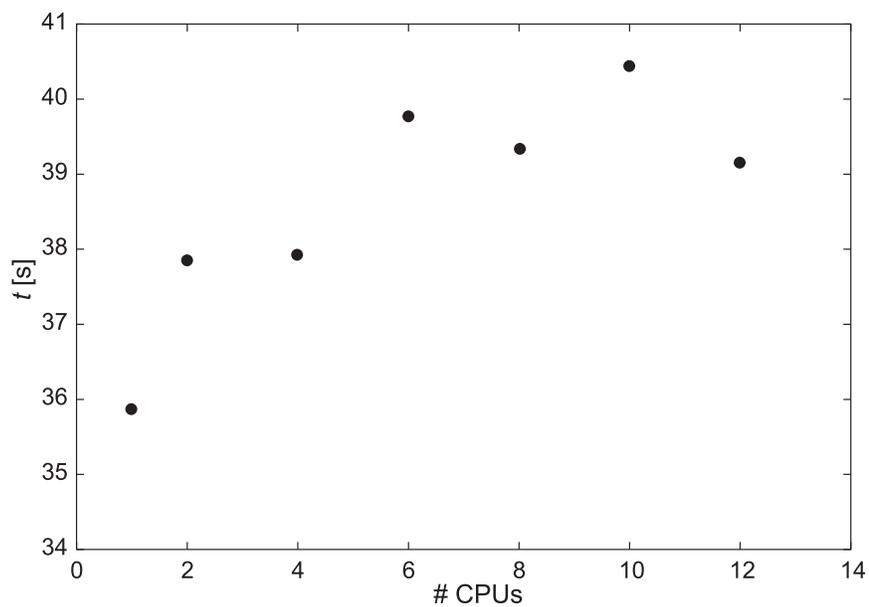}
   \caption{Scaling of the average CPU time with the
            number of available processors for the Chebyshev expansion approach
            and the LU factorization algorithm (N=500).}
   \label{fig:scale_LU}
\end{figure}

The LU factorization algorithm is an algorithm of order $N^3$ and
consequently, most of the CPU time is spent solving the linear
system of equations (see Fig.~\ref{fig:time_LU}). As a
consequence, a parallel implementation of the LU algorithm is very
difficult. Figure~\ref{fig:scale_LU} shows how the average CPU
time changes with the available number of processors. Here we use
a very simple MPI implementation of the LU algorithm as presented
in reference~\cite{ref:ibm}. Even though we could certainly
achieve better performance by employing a sophisticated LU
equation solver, the results are typical. Since the actual size of
the matrices involved is small, the communication overhead is
overwhelming and the execution time does not scale with the number
of processors.

\begin{figure}[h!]
   \centering
   \includegraphics[width=4.5in]{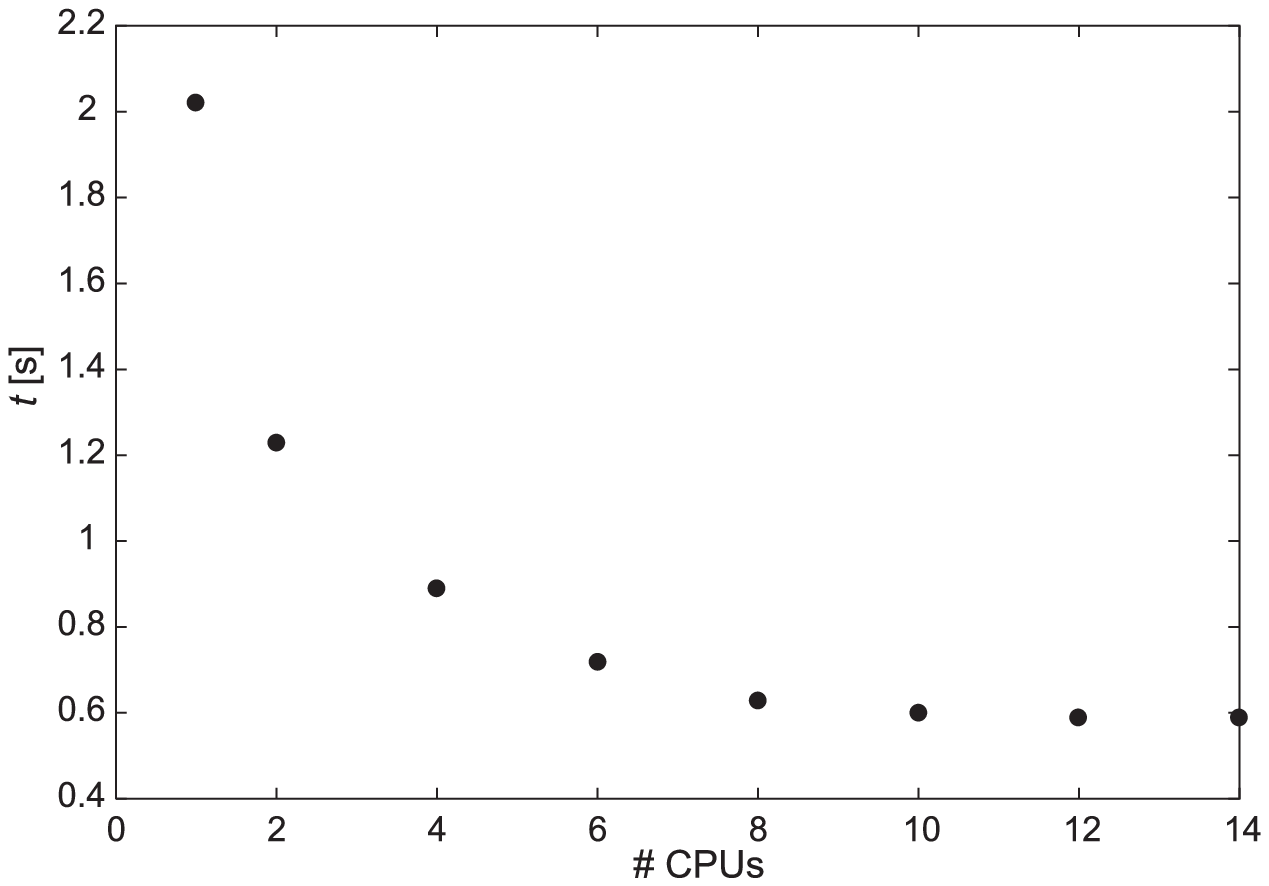}
   \caption{Scaling of the average CPU time with the
            number of available processors for the Chebyshev expansion approach
            and the biconjugate gradient algorithm (N=500).}
   \label{fig:scale_cg}
\end{figure}

Fortunately, even for dense matrices and small values of the
number of grid points~$N$, one can achieve a good parallel
efficiency. By employing an iterative method such as the iterative
biconjugate gradient method, one can render the time required to
solve the system of linear equations negligible compared with the
time required to initialize the relevant matrices, which in turn
is only slightly more expensive than the initialization process of
the LU factorization algorithm. The initialization process can be
parallelized using the algorithm presented above and the results
are depicted in Fig.~\ref{fig:scale_cg}.

It appears that by using the biconjugate gradient method the
efficiency of the parallel code has improved considerably.
However, the average CPU time saturates to give an overall speedup
of 3.5 . This can be understood by analyzing the computation and
communication requirements for our particular problem.
The calculation cost to initialize the matrices $A$ and $C$
is roughly given by the number of floating-point multiplications
and additions $(7 N^2 + 3 N) T_{\textrm {calc}}$, while the
communication cost is given by $(N^2 + 2 N) T_{\textrm {comm}}$.
Therefore, the ratio of communication to computation is
\[ \frac{N^2 + 2 N}{7 N^2 + 3N} \frac{T_{\textrm {comm}}}{T_{\textrm {calc}}}
\>.
\]
As in the finite-difference case, this ratio
approaches a \emph{constant} value as $N$ gets larger and it
becomes apparent that the communication overhead is still a
problem.

However, multi-dimensional applications such as those presented
in~\cite{ref:MDC} require complicated matrix element calculation.
In such cases, the process of initializing the matrices $A$ and
$C$ is quite involved, and the ratio of the communication time
relative to the computation time becomes favorable. 
In addition, the matrix $A$ becomes sparse and the size of the
linear system of equations is substantially larger, thus one can
also take advantage of existing parallel implementation of the
iterative biconjugate gradient algorithm~\cite{ref:pim}. Such
problems benefit heavily from an adequate parallelization of the
code. We will discuss such an example in the following section.

%
%

\section{Volterra-like integral equations for a two-point
Green function} \label{sec:ctp}

\begin{figure}[h!]
   \centering
   \includegraphics[width=3.5in]{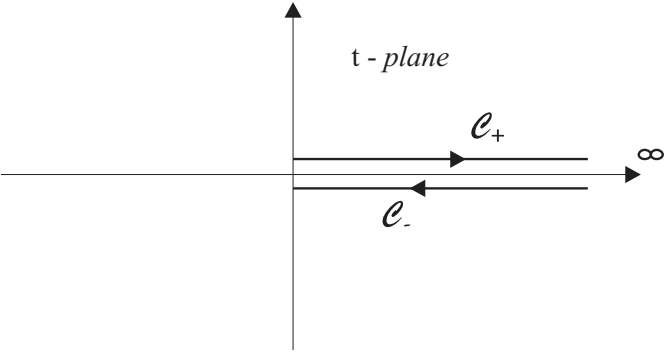}
   \caption{Complex time contour $\C$ for the closed time path
            integrals.}
   \label{fig:CTP}
\end{figure}

Schwinger, Bakshi, Mahanthappa, and Keldysh~\cite{ref:SBMK} have
established how to formulate an initial value problem in quantum
field theory. The formalism is based on a generating functional,
and the evolution of the density matrix requires both a forward
evolution from zero to $t$ and a backward one from $t$ to zero.
This involves~\cite{ref:CDHKMS} both positive and negative time
ordered operators in the evolution of the observable operators and
the introduction of two currents into the path integral for the
generating functional. Time integrals are then replaced by
integrals along the closed time path (CTP) in the complex time
plane shown in Fig.~\ref{fig:CTP}. We have
\begin{equation}
   \int_{\C} F(t) \, \mathrm{d}t \ = \
      \int_{0:\C_{+}}^{\infty} F_{+}(t) \, \mathrm{d}t -
      \int_{0:\C_{-}}^{\infty} F_{-}(t) \, \mathrm{d}t  \>.
\end{equation}
Using the CTP contour, the full closed time path Green function
for the two point functions is:
\begin{displaymath}
   \G(t,t') \ = \
      \G_>(t,t') \, \Theta_{\C}(t,t') +
      \G_{<}(t,t') \, \Theta_{\C}(t',t)  \>,
\end{displaymath}
in terms of the Wightman functions, $\G_{>,<}(t,t')$, where the
CTP step function $\Theta_{\C}(t,t')$ is defined by:
\begin{equation}
   \Theta_{\C}(t,t') =
   \left \{
      \begin{array}{ll}
         \Theta(t,t') &
            \mbox{for $t$ on $\C_{+}$ and $t'$ on $\C_{+}$} \>, \\
         0            &
            \mbox{for $t$ on $\C_{+}$ and $t'$ on $\C_{-}$} \>, \\
         1            &
            \mbox{for $t$ on $\C_{-}$ and $t'$ on $\C_{+}$} \>, \\
         \Theta(t',t) &
            \mbox{for $t$ on $\C_{-}$ and $t'$ on $\C_{-}$} \>.
      \end{array}
   \right .
\end{equation}
For complete details of this formalism and various applications,
we refer the reader to the original
literature~\cite{ref:SBMK,ref:CDHKMS}, and we confine ourselves to
discussing how our Chebyshev-expansion approach is applied to the
computation of the two-point Green function.

For simplicity we consider now the quantum mechanical limit of
quantum field theory (0+1 dimensions). In this limit, we are
generally faced with the problem of numerically finding the
solution of equation
\begin{equation}
   \G(t,t') \, = \, G(t,t') \ - \
     \int_{\C} \mathrm{d}t'' \, Q(t,t'') \, \G(t'',t')
   \>,
\label{eq:DQeqn}
\end{equation}
Here, the Green functions, $\G(t,t')$ and $G(t,t')$, are symmetric
in the sense that $ \G_>(t,t') = \G_{<}(t',t) $, and obey the
additional condition
\begin{equation}
   \G_{>,<}(t,t') \ = \
       - \ \G^{\ast}_{<,>}(t,t') \ = \ \G_{<,>}(t',t)
   \>.
\label{eq:Asym}
\end{equation}
The function $Q(t,t')$ obeys less stringent symmetries
\begin{equation}
   Q_{>,<}(t,t') \ = \ - \ Q^{\ast}_{<,>}(t,t') \ \neq \ Q_{<,>}(t',t)
   \>,
\label{eq:Qsym}
\end{equation}
which is always the case when $Q(t,t')$ has the form
\begin{equation}
   Q(t,t') \ = \
     \int_{\C} \mathrm{d}t'' \, A(t,t'') \, B(t'',t')
   \>,
\label{eq:dQ}
\end{equation}
where $A(t,t')$ and $B(t,t')$ satisfy~(\ref{eq:Asym}).

We can further write Eq.~(\ref{eq:Asym}) as
\begin{eqnarray}
   \real \{ \G_>(t,t') \} & = &
          - \ \real \{ \G_{<}(t,t') \}
   \>,
\label{eq:Are}
   \\
   \imag \{ \G_>(t,t') \} & = &
          \imag \{ \G_{<}(t,t') \}
   \>,
\label{eq:Aim}
\end{eqnarray}
or
\begin{eqnarray}
   \G_>(t,t') - \G_{<}^{\ast}(t,t')
   & = & 2 \, \real \{ \G_>(t,t') \}
   \>,
\label{eq:Adif}
   \\
   \G_>(t,t') + \G_{<}^{\ast}(t,t')
   & = & 2 \, \imag \{ \G_>(t,t') \}
   \>.
\label{eq:Asum}
\end{eqnarray}
Hence, a Green function $\G(t,t')$ is fully determined by the
component $\G_>(t,t') = \real \{ \G_>(t,t') \} + i \ \imag \{
\G_>(t,t') \}$, with $t' \leq t$. Thus, in order to obtain the
solution of Eq. (\ref{eq:DQeqn}), we only need to solve
\begin{eqnarray}
   \G_>(t,t') & = & G_>(t,t')
     - 2 \ \int_{0}^{t} \mathrm{d}t'' \,
     \real \{ Q_>(t,t'') \} \ \G_>(t'',t')
   \nonumber \\ && \quad
     + 2 \ \int_{0}^{t'} \mathrm{d}t'' \,
     Q_>(t,t'') \ \real \{ \G_>(t'',t') \}
   \>.
\label{eq:DQbig0}
\end{eqnarray}
We separate the real and the imaginary part of (\ref{eq:DQbig0})
and obtain the system of integral equations
\begin{eqnarray}
   \real \{ \G_>(t,t') \}
   & = &
   \real \{ G_>(t,t') \}
   - 2 \int_{0}^{t} \mathrm{d}t'' \,
                           \real \{ Q_>(t,t'') \}
                           \real \{ \G_>(t'',t') \}
   \nonumber \\ &&
   + 2 \int_{0}^{t'} \mathrm{d}t'' \,
                            \real \{ Q_>(t,t'') \}
                            \real \{ \G_>(t'',t') \}
\label{eq:ReDQbig0}
   \\
   \imag \{ \G_>(t,t') \}
   & = &
   \imag \{ G_>(t,t') \}
   - 2 \int_{0}^{t} \mathrm{d}t'' \,
                           \real \{ Q_>(t,t'') \}
                           \imag \{ \G_>(t'',t') \}
   \nonumber \\ &&
   + 2 \int_{0}^{t'} \mathrm{d}t'' \,
                            \imag \{ Q_>(t,t'') \}
                            \real \{ \G_>(t'',t') \}
   \>.
\label{eq:ImDQbig0}
\end{eqnarray}
The above system of equations must be solved for $t' \leq t$. The
two equations are independent, which allows us to solve first for
the real part of $\G_>(t,t')$, and then use this result to derive
the imaginary part of $\G_>(t,t')$.

Despite their somewhat unusual form, the above equations are
two-dimensional Volterra-like integral equations and our general
discussion regarding the Chebyshev spectral method applies. We
will perform a multi-step implementation of the formalism. Let
\[
   t_i = t_{i_0(N-1)+i_1}\>, \qquad 1 \le i_1 \leq N \>,
\]
be the grid location corresponding to the collocation point~$i_1$
of the interval labelled~$i_0+1$. Then, the discrete correspondent
of Eq.~(\ref{eq:DQbig0}) is
\begin{eqnarray}
   \G_>(t_i,t_j)
   &=& G_>(t_i,t_j)
   \\ &&
     - \sum_{k_0=0}^{i_0-1} \sum_{k_1=1}^N [ 2 \tilde S_{N k_1} ]
     \real \{ Q_>(t_i,t_{k[=k_0(N-1) +k_1]}) \} \G_>(t_k,t_j)
   \nonumber \\ &&
     - \sum_{k_1=1}^N [ 2 \tilde S_{i_1 k_1} ]
     \real \{ Q_>(t_i,t_{k[=i_0(N-1)+k_1]}) \} \ \G_>(t_k,t_j)
   \nonumber \\ &&
     + \sum_{k_0=0}^{j_0-1} \sum_{k_1=1}^N [ 2 \tilde S_{N k_1} ]
     Q_>(t_i,t_{k[=k_0(N-1)+k_1]}) \real \{ \G_>(t_k,t_j) \}
   \nonumber \\ &&
     + \sum_{k_1=1}^N [ 2 \tilde S_{j_1 k_1} ]
     Q_>(t_i,t_{k[=j_0(N-1)+k_1]}) \real \{ \G_>(t_k,t_j) \}
   \>,
   \nonumber
\label{eq:cheby}
\end{eqnarray}
with $t_j \le t_i$.

\begin{figure}[t!]
   \centering
   \includegraphics[width=4.5in]{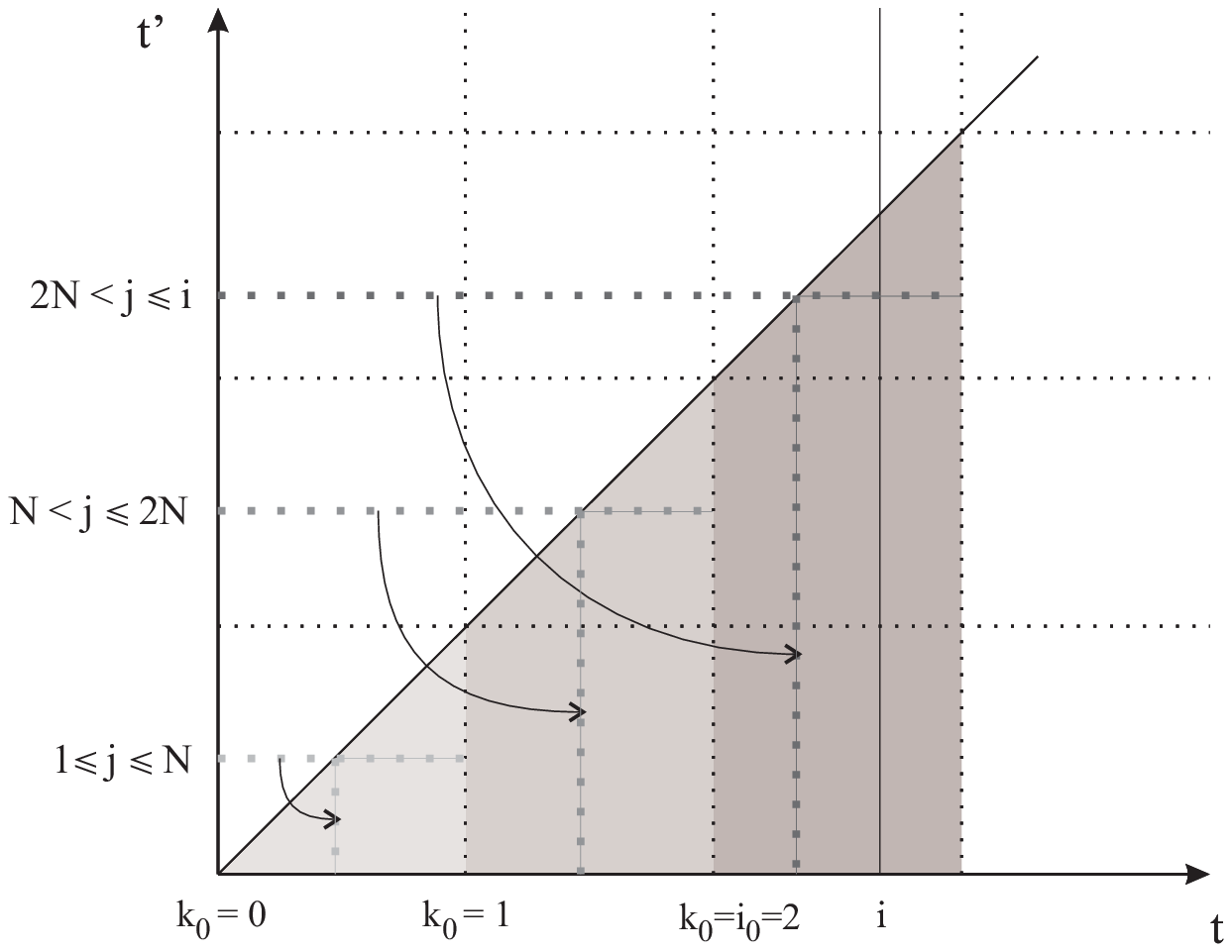}
   \caption{$\G(t_k,t_j)$ contributions to the integral $\int_0^{t_i} Q(t_i,t_k)
   \G(t_k,t_j) \textrm{d}t_k$, with $t_j \le t_i$.}
   \label{fig:suma}
\end{figure}

\begin{figure}[b!]
   \centering
   \includegraphics[width=4.5in]{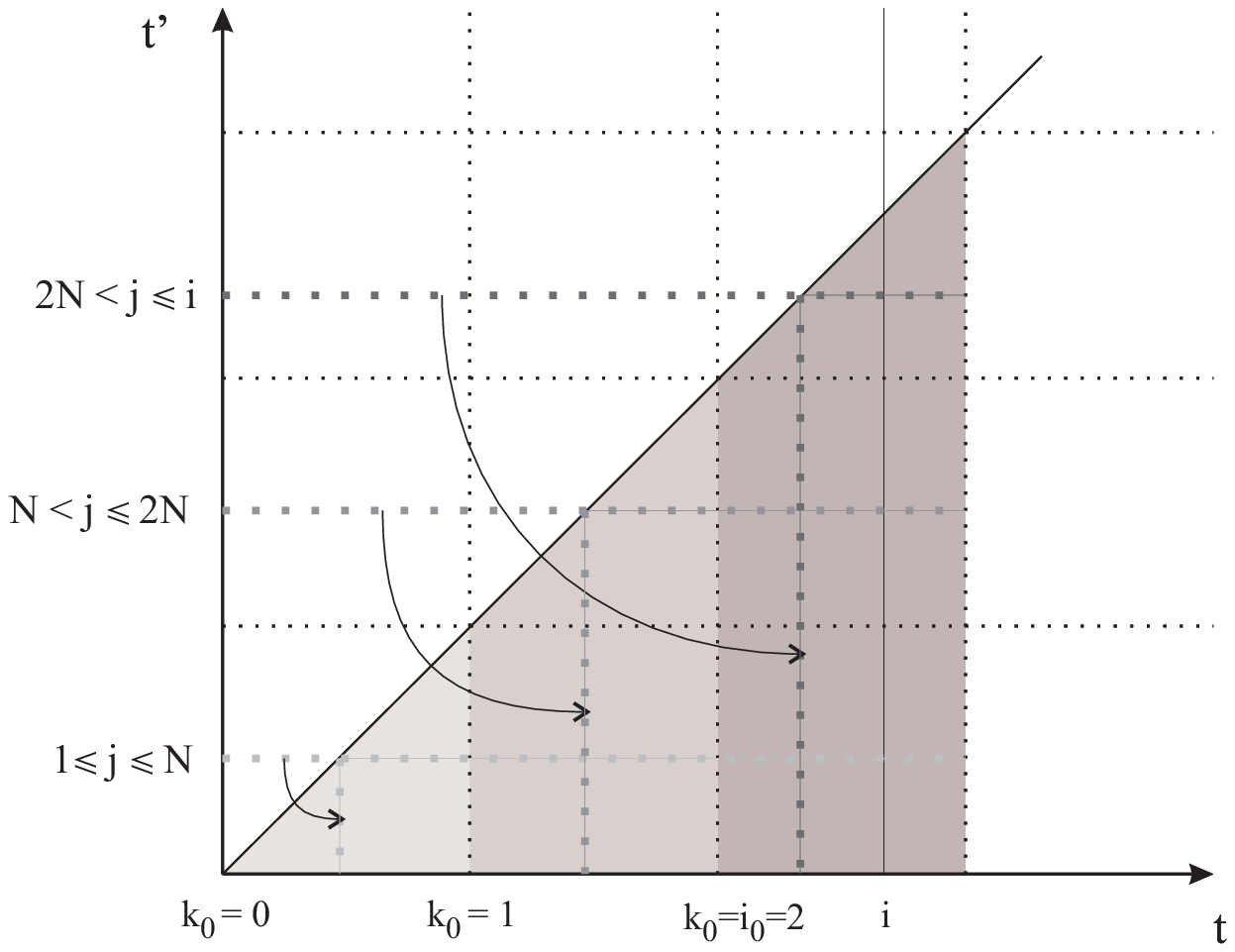}
   \caption{$\G(t_k,t_j)$ contributions to the integral $\int_0^{t_j} Q(t_i,t_k)
   \G(t_k,t_j) \textrm{d}t_k$, with $t_j \le t_i$.}
   \label{fig:sumb}
\end{figure}

We will refer now to Figs.~\ref{fig:suma} and \ref{fig:sumb}.
Equation~(\ref{eq:cheby}) involves values of $\G_>(t_k,t_j)$, for
which $t_j > t_k$. In such cases, we use the symmetry
$\G_>^\ast(t_j,t_k)$, which relates to the values the two-point
function located in the domain of interest. For the time interval
$(i_0+1)$ the size of the linear system of equations we need to
solve is
\begin{eqnarray*}
   &&
   \frac{1}{2}(i_0+1)(N-1)[ (i_0+1)(N-1) + 1 ]
   -
   \frac{1}{2}i_0(N-1)[ i_0(N-1) + 1 ]
   \\ &&
   =
   i_0 (N-1)^2 + \frac{1}{2} N(N-1)
   \>,
\end{eqnarray*}
or of order $(i_0 + \frac{1}{2}) (N-1)^2$. In practice, the value
of $N$ is taken between 16 and 32.

Tables \ref{tab:real} and \ref{tab:imag} summarize the number of
floating-point operations performed in order to compute the
non-vanishing matrix elements corresponding to a given~$i$
and~$j,\ (j<i)$.


We can now calculate the ratio of communication to computation
time, by noticing that the numbers in the tables above get
multiplied by N, corresponding to the number of collocation points
in each time step and summing over the number of steps, i.e. we
evaluate
\[
   N \Bigl [ \textrm{if}\ j > i_0 (N-1) \Bigr ]
         \ + \
   N \sum_{j_0=1}^{i_0} \
         \Bigl [ \textrm{if}\ j \le i_0 (N-1) \Bigr ]
   \>.
\]
In Table~\ref{tab:totals} we summarize all relevant estimates
regarding the computation cost for a fixed value of $i$. In order
to estimate the \emph{total} communication and computation cost,
respectively, these numbers must be multiplied by an additional
factor of $N$, corresponding to the number of possible values of
$i$ in a time step. This factor is not relevant for estimating the
communication overhead, but it must be remembered when one infers
the sparsity of the corresponding system of equations.


To conclude we observe that the communication to computation ratio
approaches
\[ \frac{1}{2(i_0+1)} \frac{T_{\textrm {comm}}}{T_{\textrm
{calc}}} \] for large values of $i_0$. Therefore for this problem
the communication overhead is reduced substantially in the later
stages of the calculation. In practice, this ratio is actually
much better, as we compute the functions $G(t,t')$ and $Q(t,t')$
on the fly, and this adds considerably to the computational
effort. Finally the sparsity of the resulting systems of equations
goes to $2/(i_0 N)$ for large values of $i_0$ and $N$, which
supports our choice for an iterative equation solver.

%
%

\section{Conclusions}
\label{sec:concl}

We have presented a numerical method suitable for solving
non-linear integral and integro-differential equations on a
massively multiprocessor machine. Our approach is essentially a
standard perturbative approach, where one calculates corrections
to an initial guess of the solution. The initial guess is designed
to satisfy the boundary conditions, and corrections are expanded
out in a complete basis of N Chebyshev polynomials on the grid of
(N+1) extrema of $T_N(x)$, the Chebyshev polynomial of first kind
of degree~N. The spectral character of the convergence of the
Chebyshev-expansion approach is the key element in keeping low the
number of grid points. From a computational point of view, each
iteration involves two stages, namely initializing the relevant
matrices and solving the linear system of equations. Both stages
can be rendered parallel in a suitable manner, and the efficiency
of the code increases when applied to complicated multi-step,
multi-dimensional problems.

The algorithm discussed in this paper represents the backbone of
current investigations of the equilibrium and nonequilibrium
properties of various phenomenological Lagrangeians. In particular
we are interested in studying the properties of the chiral phase
transition at finite density for a 2+1 dimensional four-fermion
interaction as well as the dynamics od 2-dimensional QCD, with the
ultimate goal of indirectly obtaining insights regarding the time
evolution of a quark-gluon plasma produced following a
relativistic heavy-ion collision.

\ack

The work of B.M. was supported in part by the U.S. Department of
Energy, Nuclear Physics Division, under contract No.~W-31-109-ENG-38.
The work of R.S. was supported in part by
the Natural Sciences and Engineering Research
Council of Canada under grant No.~OGP0170170.
Parallel calculations are made possible by grants
of time on the parallel computers of the Mathematics and Computer
Science Division, Argonne National Laboratory. B.M. would like to
acknowledge useful discussions with John Dawson and Fred Cooper.

%
%

%
%

%
%

\newpage

\begin{table}[h!]
   \caption{Summary regarding the calculation of $\real \G(t_i,t_j)$
            at step $i_0+1$}
   \begin{tabular}{ccccc}
      \hline
      integral &
      domain & non-zero elements & additions & multiplications \\
      \hline
      $\int_0^{t_i} \mathrm{d}t_k$ &
      $j \le i_0 (N-1)$ & N & $i_0 N$    & $(2i_0+1)N$ \\
      $\int_0^{t_j} \mathrm{d}t_k$ &
      $j \le i_0 (N-1)$ & 0 & $(j_0+1)N$ & $(2j_0+1)N$ \\
      \hline
      total &
       & N+1 & $(i_0+j_0+1)N+1$ & $2(i_0+j_0+1)N$ \\
      \hline
      $\int_0^{t_i} \mathrm{d}t_k$ &
      $j   > i_0 (N-1)$ & $(i_0+1)(N-1)+1$ & $i_0$ & $(i_0+1)N$ \\
      $\int_0^{t_j} \mathrm{d}t_k$ &
      $j   > i_0 (N-1)$ & $(i_0+1)(N-1)+1$ & $i_0$ & $(i_0+1)N$ \\
      \hline
      total &
       & $(i_0+1)(N-1)+2$ & $(i_0+1)(N+1)$ & $2(i_0+1)N$ \\
      \hline
   \end{tabular}
   \label{tab:real}
\end{table}

\begin{table}[h!]
   \caption{Summary regarding the calculation of $\imag \G(t_i,t_j)$ at step $i_0+1$}
   \begin{tabular}{ccccc}
      \hline
      integral &
      domain & non-zero elements & additions & multiplications \\
      \hline
      $\int_0^{t_i} \mathrm{d}t_k$ &
      $j \le i_0 (N-1)$ & N & $i_0 N$    & $(2i_0+1) N$ \\
      $\int_0^{t_j} \mathrm{d}t_k$ &
      $j \le i_0 (N-1)$ & 0 & $(j_0+1)N$ & $(2j_0+1)N$ \\
      \hline
      total &
       & N+1 & $(i_0+j_0+1)N+1$ & $2(i_0+j_0+1)N$ \\
      \hline
      $\int_0^{t_i} \mathrm{d}t_k$ &
      $j   > i_0 (N-1)$ & $(i_0+1)(N-1)+1$ & $i_0$ & $(i_0+1)N$ \\
      $\int_0^{t_j} \mathrm{d}t_k$ &
      $j   > i_0 (N-1)$ & 0 & $(i_0+1)N$ & $(2i_0+1)N$ \\
      \hline
      total &
       & $(i_0+1)(N-1)+2$ & $(i_0+1)(N+1)$ & $(3i_0+2)N$ \\
      \hline
   \end{tabular}
   \label{tab:imag}
\end{table}

\begin{table}[h!L]
   \caption{Global communication and computation data regarding the calculation
            of $\G(t_i,t_j)$ at step $i_0+1$}
   \begin{tabular}{ccc}
      \hline
      equation & floating-point numbers to be sent & floating-point operations \\
      \hline
      $\real \G(t,t')$ &
      $(3i_0+1.5)N^2 - (3i_0 + 0.5) N$ & $(5.5 i_0+2)(i_0+1) N^2 + Ni_0$
                                        \\
      $\imag \G(t,t')$ &
      $(3i_0+1.5)N^2 - (3 i_0 + 0.5) N$ & $[(5.5 i_0+2)(i_0+1)+i_0] N^2 + Ni_0$
                                        \\
      \hline
   \end{tabular}
   \label{tab:totals}
\end{table}

%
%

\newpage

\end{document}